\documentclass[fleqn,usenatbib]{mnras}
\usepackage{txfonts}

\usepackage[T1]{fontenc}
\usepackage{ae,aecompl}
\usepackage{graphicx}	
\usepackage{amsmath}	
\usepackage{amssymb}	
\usepackage{mathtools}
\usepackage{subcaption}
\usepackage[dvipsnames]{xcolor}
\usepackage{multirow}
\usepackage{pdfpages}
\usepackage{stfloats}
\usepackage{tabularx}
\usepackage{float}
\usepackage[export]{adjustbox}
\restylefloat{table}

\newcommand{\be}{\begin{equation}}
\newcommand{\ee}{\end{equation}}

\newcommand{\bt}[1]{{\textcolor{Plum}{#1}}}

\newcommand{\order}[1]{\mathcal{O}\!\left(#1\right)}

\title[CS detection by ML]{{\it Planck} Limits on  Cosmic String Tension Using Machine Learning}
\author[Torki et al.]{
M. Torki$^1$\thanks{M. Torki and H. Hajizadeh contributed equally to this work as first authors.},
H. Hajizadeh$^1$\footnotemark[1],
M. Farhang$^1$\thanks{corresponding author: m\_farhang$@$sbu.ac.ir},
A. Vafaei Sadr$^{2,3}$ \&
S. M. S. Movahed$^{1}$
\\
$^{1}$Department of Physics, Shahid Beheshti University,  1983969411, Tehran, Iran\\
$^{2}$Department de Physique Theorique and Center for Astroparticle Physics, University Geneva, 1211 Geneva, Switzerland\\
$^{3}$Institute for Research in Fundamental Sciences (IPM), P. O. Box 19395-5531, Tehran, Iran
}
\pubyear{2021}

\begin{document}
\label{firstpage}
\pagerange{\pageref{firstpage}--\pageref{lastpage}}
\maketitle

\begin{abstract}
We develop two parallel machine-learning pipelines to estimate the contribution of cosmic strings (CSs), conveniently encoded in  their tension ($G\mu$), to the anisotropies of the cosmic microwave background radiation observed by {\it Planck}. 
The first approach is tree-based and feeds on certain map features derived by image processing and statistical tools. The second uses convolutional neural network with the goal to explore possible non-trivial features of the CS imprints. 
The two pipelines are trained on {\it Planck} simulations and when applied to {\it Planck} \texttt{SMICA} map yield the $3\sigma$ upper bound of $G\mu\lesssim 8.6\times 10^{-7}$. 
We also train and apply the pipelines to make forecasts for futuristic CMB-S4-like surveys and conservatively find their minimum detectable tension to be $G\mu_{\rm min}\sim 1.9\times 10^{-7}$.
\end{abstract}



\section{Introduction}
Cosmic strings (CSs) are hypothetical one-dimensional topological defects possibly formed in certain cosmic phase transitions in the early Universe \citep{Kibble:1976sj,Zeldovich:1980gh,Vilenkin:1981iu,
Vachaspati:1984dz,Vilenkin:1984ib,Shellard:1987bv,
Hindmarsh:1994re,Vilenkin:2000jqa,Sakellariadou:2006qs,
Bevis:2007gh,Depies:2009im,Bevis:2010gj,
Copeland:1994vg,Sakellariadou:1997zt,Sarangi:2002yt,
Copeland:2003bj,Pogosian:2003mz,Majumdar:2002hy,
Dvali:2003zj,Kibble:2004hq,HenryTye:2006uv}.
Their detection would give a unique opportunity to study the high energy physics of the relevant phase transition. 
The CS network, if existing, can source inhomogeneities in the distribution of matter and background photons. 
Among the different CS imprints is the Gott-Kaiser-Stebbins (KS) effect \citep{Kaiser:1984iv}. This effect describes the line-like discontinuities left by moving strings on the Cosmic Microwave Background radiation (CMB) temperature  through the integrated Sachs-Wolfe effect
\citep{lazanu2015constraints,Gott:1985,Stebbins:1988,bouchet1988microwave,allen1997cmb,pen1997power,ringeval2012all}.
The induced anisotropy has the form \citep{Hindmarsh:1993pu,Stebbins:1995}
\begin{equation}
\frac{\delta T}{T} \sim 8 \pi G\mu v_{\rm s}
\end{equation}
where $G\mu$ is the main parameter characterizing the string network with  $\mu$ being the string tension and  $v_{\rm s}$ the transverse velocity of the string. 
The string tension  is intimately related to 
the energy scale of the phase transition through 
\begin{equation}
\frac{G\mu}{c^2}=\order{\frac{\varpi^2}{M_{\rm Planck}^2}},
\end{equation}
where   $\varpi$ represents the energy scale of the transition. The constants $G$  and $c$ are the Newton constant and the speed of light and $M_{\rm Planck}\equiv\sqrt{\hbar c/G}$ is the Planck mass. 

There have been a lot of efforts to model and simulate these imprints and various statistical tools have been developed for their study.
In particular, CMB-based studies using {\it Planck} temperature and polarization power spectra yield 
 $G\mu < 2.0\times 10^{-7}$ for Higgs strings (Lizarraga et al. 2016), $G\mu < 1.5\times 10^{-7}$  for  Nambu-Goto strings (Lazanu  Shellard 2015) and  $G\mu < 1.1\times 10^{-7}$  for a multi-parameter fit to the unconnected segment model \citep{cha16}. 
Methods based on the non-Gaussianity of the string-induced anisotropies  \citep{rin10,duc12} such as bispectrum and Minkowski functional measurements and wavelet analysis  give 
$G\mu < 8.8\times10^{-7}$, $ G\mu < 7.8\times10^{-7}$ and $G\mu < 7\times10^{-7}$ respectively \citep{hin09,hin10,reg15,ade2014planck,duc12}.
Map-based approaches use real-space statistics  or  edge-like properties of the KS imprints to tighten the detectability limit of the strings.  
For instance, there are detectability claims from  ideal noise-free simulations for CSs with $G\mu > 4.0\times 10^{-9}$  using crossing statistics \citep{mov10}
and with  $G\mu > 1.2 \times 10^{-8} $ with the unweighted Two-Point Correlation Function (TPCF) of CMB peaks \citep{mov12}. A recent analysis of the peak clustering  provided the upper bound of  $G\mu\lesssim 5.59\times 10^{-7}$  in \texttt{SMICA} \citep{vafaei2021clustering}.

%
For simulations of the South Pole Telescope, \cite{ste08} set the minimum detectability bound of  $G\mu >  5.5\times10^{-8}$ with edge-detection algorithms and  \cite{her16} find a detection limit of $G\mu >  1.4\times10^{-7}$ using wavelets and curvelets. 
\cite{vaf17} also used various combinations of image processing tools and statistical measures,  followed by tree-based learning algorithms \citep{vaf18}  to forecast the detectability limit of different observational scenarios. 
There have been relatively recent neural network-based algorithms to 
locate the position of the strings for ideal noiseless experiments \citep{ciu17} or to put information-theoretic bounds on the CS tension for various observational setups \citep{ciu20}.
There are  also $G\mu$ measurements and forecasts from other observational routes. For example see \cite{rin17,bla17a,bla17b,arz18,bla18,auc19,len15,jen06,psh09,tun10,dam05,bat10,kur12} for bounds on string loops from pulsar timing constraints on the amplitude of the stochastic gravitational wave background, \cite{imt20} for bounds on CS network from fast radio bursts, \cite{bra10,her12,pag12} for forecasts from  21-cm signatures, \cite{lal19} for CS imprints on ionization fraction in the Universe, \cite{fer20} for the effect of CS on the filament structure in the cosmic web and \cite{cun18} for the imprint of CSs on the dark matter distribution.

In this manuscript, we propose to use convolutional neural networks (CNN) to search for the CS footprints on CMB data. 
Deep learning methods, i.e. neural networks containing many layers, have led to significant improvements in performance over the past decade in almost all machine learning problems. 
Here, we study the possibility of achieving tighter upper bounds on the CS tension by deep learning methods from {\it Planck}18 data  and compare the results with the limits from an improved version of the previous approaches of \cite{vaf17,vaf18}.
%
%
We also apply the same two pipelines, properly trained, to other observational setups to compare their performance and make forecast for future surveys. 

The outline of the rest of this paper is as follows. 
Section~\ref{sec:ml} is a brief introduction to the machine learning algorithms used. An overview of the CMB simulations and data  is given in Section~\ref{sec:sim}. In Section~\ref{sec:det} the methodology and pipelines are introduced and discussed in detail. In Section~\ref{sec:res} we discuss and compare the performance of the models and apply them to {\it Planck}18 data. We conclude with a summary and a short discussion in Section~\ref{sec:sum}.

\section{Machine Learning}\label{sec:ml}

Machine learning is the branch of artificial intelligence with the goal of empowering computers to learn and improve from experience. 
In machine learning, given some data, a machine is trained to construct a model which is used to make predictions on new datasets. 
If supervised, the target value of the learning algorithm is provided by an instructor. In unsupervised learning, on the other hand, the algorithm learns to make sense of  data with no guide. This happens by experiencing a dataset with many features which are not pre-labeled and trying to learn its inherent structure, usually through clustering \citep{goodfellow2016deep}.

In the rest of this section we briefly review the two main learning algorithms used in this paper, namely the gradient boosting method and neural networks.
\subsection{Gradient Boosting Method}
Gradient boosting models are supervised algorithms based on decision trees and are commonly used in classification and regression problems. 
A decision tree has a tree-like structure with nodes, branches and leaves.
The tree starts with a question, corresponding to a node, and depending on the answer, the data are split into branches. The process continues until the algorithm reaches a leaf that predicts the label of the data.
Gradient boosting combines many weak learners into a strong learner and is based, not on a single model, but on gradually improving the model toward models with lower loss function and higher prediction power at each step.
The performance of a model is evaluated by the loss or cost function through  
comparing predictions with fiducial values. 

In this work we use the two gradient boosting libraries of XGBoost\footnote{https://xgboost.readthedocs.io/en/latest/} \citep{che16} and LightGBM\footnote{https://lightgbm.readthedocs.io/en/latest/} \citep{lig17}, or LGBM for short. 
XGBoost is an advanced and powerful implementation of gradient boosting and due to its high accuracy, is the popular boosting method for many learning problems.  
However,  XGBoost can be very time-consuming  for  problems with huge datasets. 
In LightGBM, the tree grows leaf-wise in contrast to XGBoost and other tree-based algorithms where  the tree grows depth-wise. The leaf-wise split surprisingly increases the speed and accuracy of the model.
However, due to increased complexity, there is a high risk of being overfit. This is usually avoided by setting a maximum depth to which splitting can occur. 
The leaf with a larger loss function is chosen to grow and  further reduce  loss. 
It is therefore faster, the reason it is called {\it light}.
\subsection{Neural Networks}

Artificial Neural Networks (ANNs), inspired by biological neural networks,  consist of neurons with cell body and dendrites to receive inputs and axons to output a signal \citep{jain1996artificial,hassoun1995fundamentals,yegnanarayana2009artificial}. The neural network takes the data in its first layer and calculates a weighted sum over them.  The weights are initialized  randomly, and are updated  during the training process.
  The higher the assigned weight to an input, the more significant it has been considered compared to other features. Through learning,  the model finds weights that result in the desired behavior of  the network. 
    A bias is usually added to the weighted sum of the input to change the output range and lead to predictions with better fit to data.
 A non-linear differentiable transformation is applied to weighted data by an activation function. This step increases the complexity of the model and without it the model would be a simple linear regressor.
Among the activation functions commonly used are sigmoids (commonly $\frac{1}{1+e^{-x}}$), softmax functions or $\frac{e^{x_i}}{\sum_j{e^{x_j}}}$,  rectifiers or ${\rm max}(0,x)$ and $\tanh(x)$ \citep{nwankpa2018activation}.
The output of a neuron is fed into the next layer and these steps are repeated until combined into the final layer where a prediction is made.
The algorithm is called deep learning if many hidden layers are used with long causal chains of computational steps. 
The learning process is basically finding the parameters  of the network
that optimize the cost function. 
In each iteration the learning rate (LR) controls the weight update with respect to the loss gradient,  new weights = old weights - LR $\times$ gradient of loss. LR is often a small number between 0 and 1. A good idea is to start training with a large LR and gradually decrease it in the training process.

Convolutional neural networks (ConvNet or CNN) are deep learning algorithms
 that use convolution instead of general matrix multiplication in at least one of their layers \citep{krizhevsky2012imagenet,gu2018recent}.
Convolution can be interpreted as a filter, kernel or weight matrix that multiplies part of the input data 
to produce a convolved feature map.
Neurons that lie in the same feature map share the weight.   
The network complexity is thereby reduced by keeping the number of parameters low \citep{hinton2012improving}.
Similar to weights in regular neural networks, the filters are initiated randomly but 
are updated in the training process to optimize the cost value.
Different hyperparameters  that control the output size include 1- number of filters at a layer that specifies the output depth of a convolution layer, 2- kernel-size that specifies the dimension of the convolution matrix and controls the width and height of output, 3- stride or step of filter movement, and 4- zero-padding  that adds extra layers of zeros across the images to keep the output size the same as the input \citep{aloysius2017review}.

CNNs may also contain {\it pooling} layers that reduce the dimension of the data by combining  the outputs of certain neurons of a layer into a single 
neuron in the next layer. 
This is achieved by, e.g., taking the maximum output of the neuron cluster in the pooling region (max pooling, see \cite{cir12}) or by computing their average (average pooling, see \cite{mit20}). 
Pooling would reduce the computational complexity of the model.  

A well-trained machine should easily generalize to unseen (test) datasets in the sense that it has small errors when making predictions on them. A gap can occur between the predicted errors of train and test datasets  if the model is overfit by learning insignificant and noisy  features of the train sets and having partially missed the main underlying structure of data. {\it Dropout} is one of the techniques proposed to prevent overfitting by randomly and temporarily removing  a unit from the network \cite{sri14}. This technique forces the machine to learn about more robust features.  

{\it Batch Normalization} (BN) is another technique to make the model more generalizable \citep{iof15}.
In BN the model is trained on data chunks with equal size, or batches, instead of the whole dataset in a single go. The BN layer increases the learning rate  and reduces the sensitivity to initialization. 
\section{Simulations and Data}
\label{sec:sim}
In this work, we use simulations to test, train and verify the robustness of the proposed CS search pipelines.
The simulated maps contain contributions from inflationary Gaussian CMB anisotropies and different levels of CS-induced fluctuations and also include experimental effects.
We use Nambo-Gutto string  simulations in this work \citep{bennett1990high,ringeval2007cosmological}.

The  simulation for the observed CMB sky map, $d$, would then be
\begin{equation}
\label{eq:d}
d = T+ B \left(G\mu \times S\right) + n
\end{equation}
 where $S$, $n$ and $B$ represent the template for the CS-induced anisotropies, the instrumental noise and beam respectively. 
 The simulated anisotropies $T$ represent the observed inflationary Guassian fluctuations and in this work in divided into three main sets. 
 %
%
The first set consists of HEALPix\footnote{https://healpix.sourceforge.io}-generated CMB maps, $g$, with $T=Bg$ in Equation~\ref{eq:d}.
This is used in this work with simulations of the two phases of CMB-S4, here referred to as CMB-S4-like(I) and (II).  
The second and third sets of simulations are {\it Planck}18 simulations, labeled by the Full Focal Plane 10\footnote{pla.esac.esa.int $\to$ maps $\to$ simulations $\to$ cmb} (FFP10)  and the end-to-end\footnote{pla.esac.esa.int $\to$ advanced\;search\;and\;map\;operations $\to$ simulated\;maps\;search $\to$ comp-separation} (E2E) simulations \citep{ade2016planck}.
The FFP10 maps are generated using the best characterizations of instrumental properties, as well as astrophysical foreground residuals. 
The instrumental noise, $n$, for all these cases is assumed  Gaussian and white  with the desired noise level. 
 For the details of experimental characterization for the simulations used in this work see \cite{vaf18}. 
The E2E maps are constructed based on the FFP10 simulations, where these maps go through the same pipeline as the data \citep{akrami2019planck,ade2016planck}. The E2E CMB and noise  maps, $T$ and $n$ respectively, are therefore the closest to observations. %
 
\begin{figure*}
	\centering	
	\includegraphics[width=\textwidth]{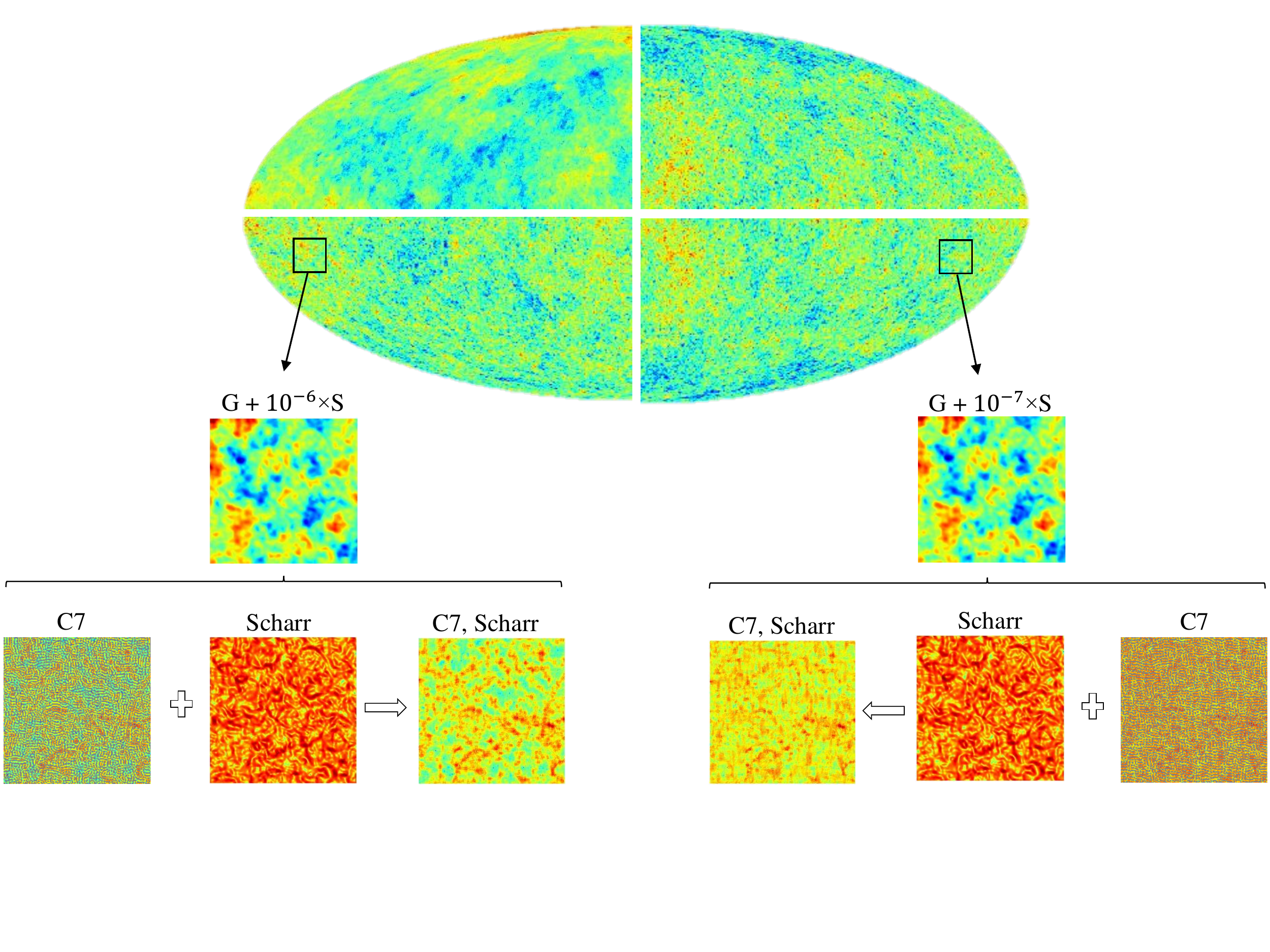}
	\caption{Up: The full sky map, with $N_{\rm side}=2048$, divided into four patches to represent different simulations. The upper half correspond to E2E (right) and CS-induced (left) simulations. The lower parts show the sum of the E2E and string simulations with $G\mu$ = $10^{-6}$ (left) and $G\mu$ = $10^{-7}$ (right).
	Middle: The zoomed-in view of two $256\times256$ patches with different string contributions. 
	Bottom: The same patches as the middle row, after having passed through parts of the image-processing steps of the proposed CS detection pipeline. Specifically, they correspond to the seventh curvelet component (labeled as C7), the Scharr-filtered image and the combination of the two.}
	\label{fig:CMB}
\end{figure*}

The pipelines developed based on the HEALPix simulations are used to make futuristic forecasts on CS detectability from CMB data and works with the HEALPix resolution $N_{\rm side}=4096$.
 The second and third sets of pipelines establish an estimate for the anticipated level of detectability with the {\it Planck} data, and work with  $N_{\rm side}=2048$.
The E2E maps are also used to build machine learning-based models which are then directly applied to {\it Planck}18 observations to measure or set an upper bound on $G\mu$. 
\section{detection pipelines}
\label{sec:det}

In the following, we divide the $G\mu$ range of interest into $n_{\rm cls}=11$ classes, consisting of a null case with $G\mu=0$ and ten logarithmically-scaled classes in the range $(6\times 10^{-9},6\times 10^{-6})$.  
Then $N_{\rm sim}$ simulations are generated with different levels of string contribution  for each observational case.
The two search methods are patch-based.  After removing the sky mask\footnote{pla.esac.esa.int $\to$ maps $\to$ mask $\to$ cmb $\to$ 2018\;component\;separation\;common\;mask\;in\;intensity}, the maps are divided into $N_{\rm patch}$ non-overlapping patches to reduce the computational cost. This is not  expected to lead to any noticeable  information loss on CS network
as the strings mainly leave small-scale imprints.  

To exploit the available information in the full observed sky, the results from all patches need to be interpreted simultaneously. 
For this purpose, we treat the patches as independent realizations of a single cosmological model with the same $G\mu$ and multiply their probability vectors to give the full-(masked-)sky probability vector as the final prediction of the machine.

In our search for CSs, we follow two main parallel approaches.
 The first is feature-engineered and is guided by previous experience \citep{vaf17,vafaei2021clustering}. In this approach certain features, chosen based on earlier works on the topic, are extracted from the maps 
 and fed as input to machines for classification. 
The second approach, on the other hand, uses deep learning methods to estimate string contribution to CMB anisotropies. 

\subsection{Educated Search}
\label{sec:edu}

 The CS search pipeline introduced in this section is an extension of \cite{vaf17,vafaei2021clustering} to more realistic {\it Planck} simulations of FFP10 and E2E maps to allow for direct comparison with {\it Planck} data. 
 Here the maps are processed and  compressed into features with the goal to  enhance the CS signal contribution compared to primordial CMB anisotropies. 
In this branch of the pipeline we use $N_{\rm sim}=10$ maps.
 The maps are divided into $N_{\rm patch}=192$ patches for all cases but E2E where we take $N_{\rm sim}=100$. 
That is because the machines trained on E2E maps will be applied to {\it Planck} observations, and,  as will be discussed later,  the performance of the machines  slightly improves when $N_{\rm sim}$ increases from ten to $100$. 
For the other observational scenarios, however, as no direct comparison with data is made, we use $N_{\rm sim}=10$ to keep the computation cost low.

\subsubsection{Processing maps}

In the first step, the sky patches go through two image processing steps of curvelet decomposition
\citep{don00,can00,can01,can02,can06} and Canny filtering\citep{can86},  to increase the detectability of the CS signature. 
Curvelet transformation disintegrates the maps into layers with scales relevant to the signal of interest. We use three small-scale curvelet components, labeled as C5, C6 and C7, so that only components with the highest contribution from CSs are kept for further analysis.
The curvelet decomposed maps are then filtered using Canny algorithm which is suitable for edge detection in images. In previous analysis we found the best results from Scharr and Sobel filters. These two are therefor used as filters here to produce the gradient maps.
Figure~\ref{fig:CMB} illustrates the effect of these steps on the input map, for various combinations of primordial, CS-induced and noise fluctuations.

\subsubsection{Analysis of  processed maps}
 \begin{figure}
	\includegraphics[width=\linewidth]{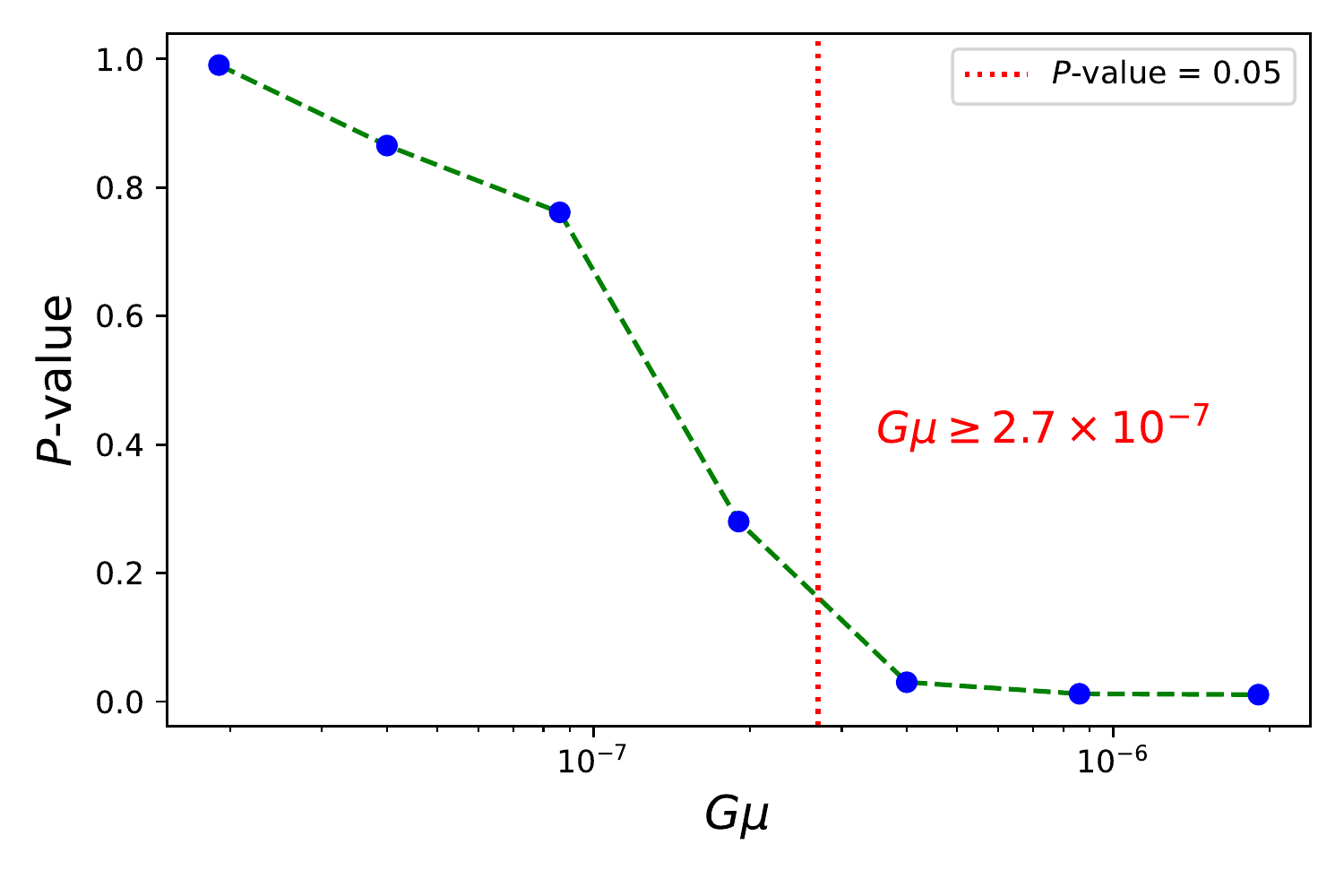}
	\caption{The {\it P}-values for the $G\mu$ classes, measured by the LightGBM method, quantify the difference between  the distributions of the predictions for each class from the null case, for the E2E simulations. The vertical line marks the minimum detectable $G\mu$   ($G\mu_{\rm min}$) where the {\it P}-value crosses the detectability threshold (taken to be $0.05$ here) and is found by interpolation.}
	\label{fig:pvallgbm}
\end{figure}
The goal of the map processing step is to enhance the detectability of the imprints of the CS network. 
Here we quantify the possible imprints by applying various statistical measures on the processed maps. 
In previous analysis of \cite{vaf17} it was shown that, among the different measures explored, 
the tightest constraints were reached by  the  standard deviation and one-point probably distribution function (hereafter, the PDF).
We thus focus on the same measures in the current work.
The standard deviation returns a number for each patch.  We subtract from this number the mean standard deviation of the  baseline patches with $G\mu = 0$.
The PDF of each patch, on the other hand, is a function. We subtract from these functions the baseline PDF and then integrate over it to associate a single number to each patch. 

At the end of the processing steps we are left with 12 features for each patch, obtained from combinations of three curvelet components (C5, C6 and C7), two filters (Scharr and Sobel) and two statistical measures (PDF and standard deviation). 
In the next section we explain how we follow a learning-based approach using these features to estimate the level of string-induced anisotropies.  
We also calculate and report  the two-tail ${\it P}$-value statistics as a traditional measure to assess  the performance of the algorithms and quantify their strength. 
With the ${\it P}$-value as the criterion, we define the detectability limit, $G\mu_{\rm min}$, as the 
minimum $G\mu$ with a distribution distinguishable from the null class with a maximum ${\it P}$-value of $0.05$. Figure~\ref{fig:pvallgbm} shows the ${\it P}$-values for different levels of $G\mu$ contribution for E2E simulations. 
For the details of the quantification of the CS-induced deviation and the corresponding {\it P}-value calculation see \cite{vaf17}.

One could also follow a purely statistical path and calculate the ${\it P}$-value to compare the distribution of the above statistical measures (pdf and variance) 
 with the null class, using $N_{\rm sim} \times N_{\rm patch}$ for different $G\mu$ classes and for all observational cases of interest. 
This is similar to the approach of \cite{vaf18}. 
Figure~\ref{fig:pvallgbm} shows the {\it P}-values of the $G\mu$s for the E2E case from this approach.
\subsubsection{Machine-learning methods}\label{sec:MLmethod}
\begin{figure}
	\centering
	\includegraphics[width=0.47\textwidth,height=0.35\textheight]{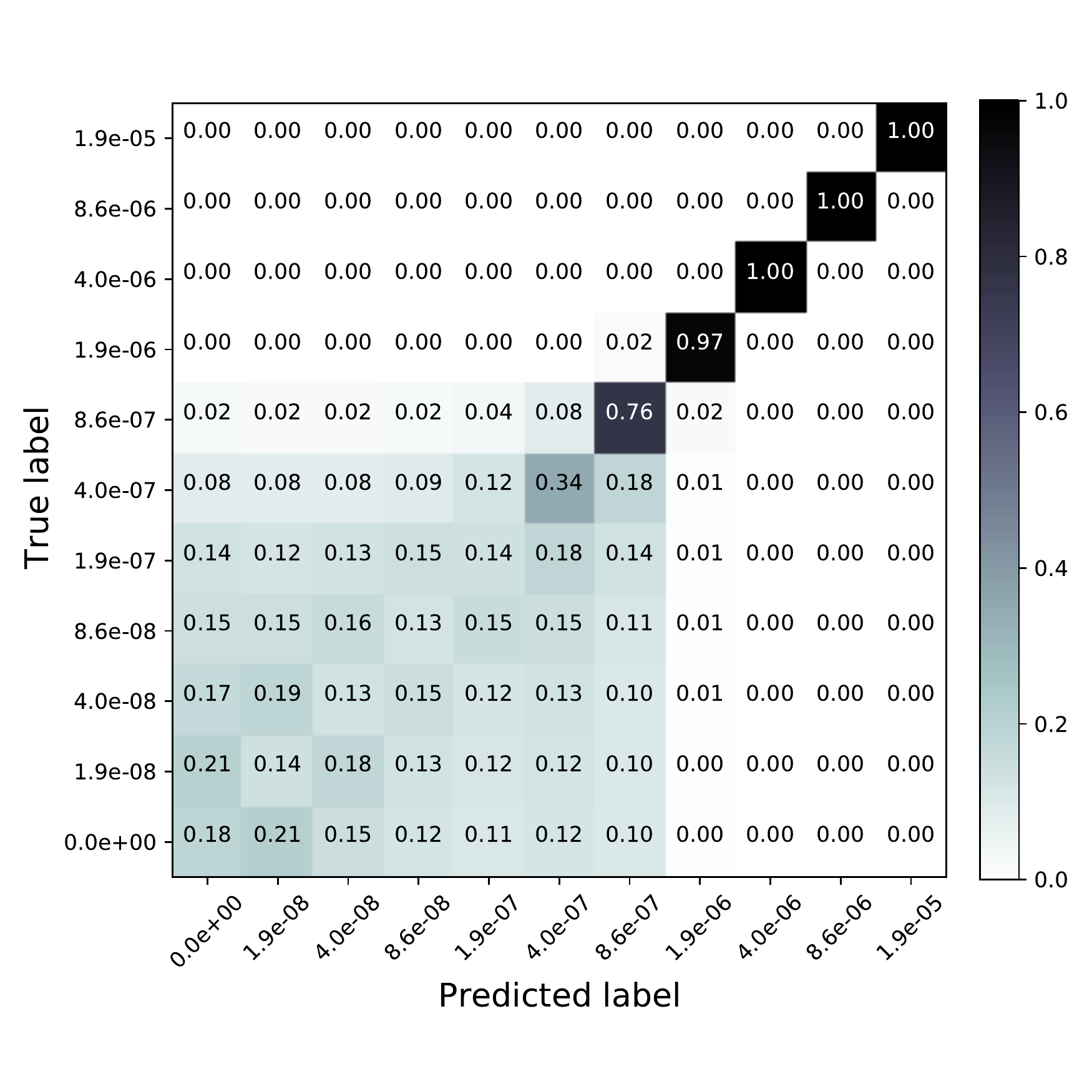}
	\caption{The confusion matrix summarizes  the results of the classification of the  LightGBM algorithm on unseen E2E simulations. The diagonal elements are the fractions of correct predictions, while the off-diagonals represent the mislabelled. We see the model has not learned much about the first six classes and the  predictions in the $6\times6$ block in the lower left are quite scattered and hardly centered around fiducial values. We, conservatively, report the first class above this confused block as the minimum detectable class by the model that can be distinguished from the null class, i.e., $G\mu_{\rm min}=8.6\times10^{-7}$.}
	\label{fig:conflgbm}
\end{figure}

We use the 12 map features of the previous section, derived from combinations of various statistical measures and image processing tools, as inputs to different tree-based learning algorithms for CS imprint detection.
For each observational scenario we use $75\%$ of the simulated patches for training  and the remaining as test sets to assess the power of the trained model.
We compare the performance of different machine-learning methods including Naive Bayes \citep{ris01}, Decision Tree \citep{qui86}, Random Forest \citep{bre01}, K-Nearest Neighbours \citep{kel85}, XGBoost \citep{che16} and LightGBM \citep{ke17} .
We found the boosting methods of XGBoost and LightGBM as the most accurate, with LightGBM being the fastest of the two. The LightGBM is therefore the main learning algorithm in this work for the tree-based search, and the reported results are based on this algorithm.
The performance of the model after training is summarized in the confusion matrix.
Figure~\ref{fig:conflgbm} shows the confusion matrix of the 11 $G\mu$ classes for the E2E simulations. The columns and rows of this matrix represent the predicted and actual classes respectively. 
We see that the upper right ($5\times5$) diagonal block of this matrix has a dominant diagonal element with diminishing values as one moves to either side. This feature is the illustration of the low confusion of the machine in learning about the signal we are after in this regime, and making relatively powerful predictions about the $G\mu$ classes. 
On the other hand, the lower left ($6\times6$)  block of the matrix and the scattered distribution of the predictions for these classes show that the machine has not learned about these classes and the predictions are not reliable. We therefore consider the lowest class above this {\it confused block} as the minimum detectable $G\mu$ or $G\mu_{\rm min}$. 
It should, however, be noted that detectability requires not only robust predictions based on powerful learning by the machine, but also distinguishability from the null class with  $G\mu=0$. 
Therefore, for a (non-zero) $G\mu$ to be chosen as the detection limit,  the majority (e.g., $95\%$) of the machine predictions for the patches are non-null. Figure~\ref{fig:conflgbm} shows that $G\mu_{\rm min}= 8.6 \times 10^{-7}$ is the detection limit for the educated method in this work based on the confusion matrix.
This is a conservative choice due to the discrete nature of the classes. One could possibly decrease this limit by zooming into the neighborhood of this $G\mu_{\rm min}$ and further train the machine on fine-gridded classes. 

The process of machine learning in this section is separate and totally different from the deep search that will follow in section \ref{Deep_sec}. 
The  features here are chosen a priori and handed in to the machine as inputs, while in the deep search, the network itself selects the best features for training. The only preprocessing used in the deep search is  edge-enhancement through filtering the map.

In the educated search
the machines are trained based on our prior assumption that certain features effectively represent or enhance the desired signal. So the maps are initially transformed into these features to improve the $G\mu$ detection limit   
of the models. 
However, there is no guarantee that  our search in the space of statistical measures and image processors is exhaustive. 
 For instance,  there could exist nonlinear features that are most sensitive to the string network and are left out in our limited and linear image processors. 
Therefore, as a parallel approach, we use CNNs to classify the input maps according to their predicted string contribution. 
The main advantage of deep learners over the educated algorithms is that the network itself learns to extract features in the training process. 
In this section we introduce our deep string detector. 

\subsubsection{Input preparation for the CNN}
We train the network with patches of $256\times 256$ pixels  ($\sim7.5\times7.5 ~  \rm{deg}^2$),  randomly chosen  from $N_{\rm sim}\sim 60$ full sky simulations with $N_{\rm side}=2048$. The string contribution in each simulation is set by a randomly chosen $G\mu$. The maps are first Scharr-filtered to enhance their edge-like features and then standardized by subtracting their mean and normalizing their standard deviation to unity. 

As for any supervised algorithm, the targets should be labeled. One could use one-hot vectors as the labels for the $n_{\rm cls}$ classes of $G\mu$, i.e., $n_{\rm cls}$ arrays with a single one and $n_{\rm cls}-1$ zeros.
 The one component in each vector represents the $G\mu$ class of the input map. 
 The  prediction for each map is also an array with 
 $n_{\rm cls}$  components between 0 and 1, specifying the probability of the data  belonging to each class.  If the targets are labeled  by their $G\mu$ values instead, the network will treat the problem as a regression. 

\begin{table}
\caption{The final architecture of CNN. The abbreviations used in this table are as follows. f: number of filters, k-size: kernel size, s: number of strides, p: pooling-size, AF: Activation Function, BN: Batch Normalization and Conv.: convolution layer.}
\label{Table:arch}
\centering
\begin{tabular}{c}
\hline
$4 \times$ \Bigg\{ Conv.(f=4 , k-size = 5 , s=1) \\
$\rightarrow$ BN \\
$\rightarrow$ 
CReLU activation layer \Bigg\} \\
$\downarrow$ \\
$6 \times$ \Bigg\{ Conv.(f=8 , k-size = 5 , s=1) \\
$\rightarrow$ BN \\
$\rightarrow$ CReLU activation layer \\
\\
$\downarrow$ \\
\\
Conv.(f=16 , k-size = 5 , s=2) \\
$\rightarrow$ BN \\
$\rightarrow$ CReLU activation layer  \\
$\rightarrow$ Max-Pooling(p=2 , s=1) \Bigg\} \\ 
$\downarrow$ \\
Flatten \\
$\downarrow$ \\
Dropout(rate = 0.5) \\
$\downarrow$ \\
Dense(40 , AF= CReLU) \\
$\downarrow$ \\
Dropout(rate = 0.5) \\
$\downarrow$ \\
Dense(20 , AF= CReLU) \\
$\downarrow$ \\
Dropout(rate = 0.5) \\
$\downarrow$ \\
Dense(11 , AF= Softmax) \\
\hline
\end{tabular}
\end{table}

\subsubsection{Network architecture and analysis}
Network design is about choosing a proper architecture and tuning its hyperparameters to control the learning process, such as the number and size of kernels in each layer and the stride values.  
We try various architectures and parametrizations in two separate and parallel paths of regression, with continuous  $G\mu$ labels, and classification,  with $n_{\rm cls}=11$ . 
The former regression problem, not surprisingly, turned out to be more challenging for the network to solve. We therefore report the classifier results.
We also find that certain features of the network have significant impact on its performance and the accuracy of its predictions. 
For example, using min-max pooling layer rather than average pooling, using 
z-score normalization as the scaling choice, assigning more strides to the convolution layer than to the pooling layer, and use of batch normalization layers greatly impact the results.

\begin{figure*}
\centering
	\includegraphics[width=\textwidth , height=0.3\textheight]{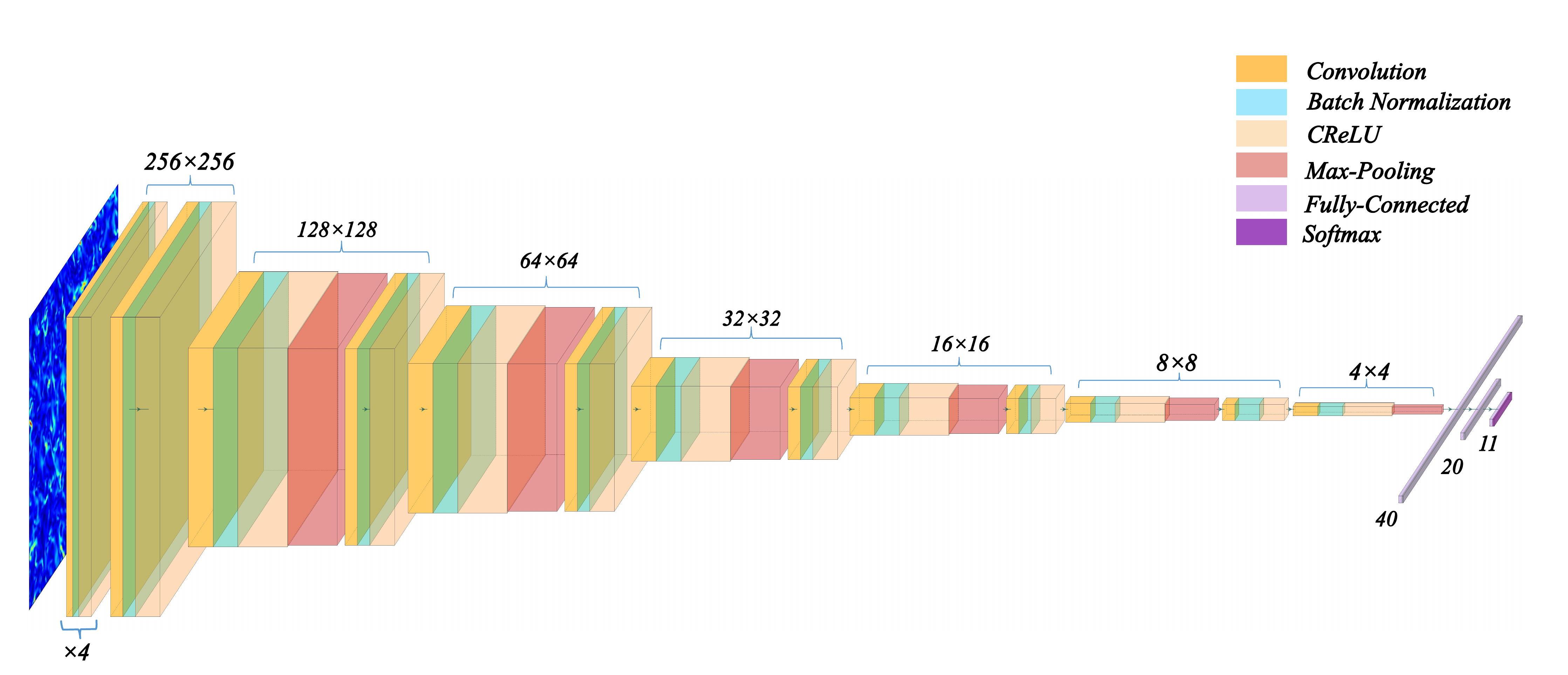}
	\caption{Visualization of the CNN architecture and variations in the input dimension through different layers (see Table~\ref{Table:arch}).  The CNN feeds on $256\times256$ images. Dimension reduction in the network results in 1D arrays with $n_{\rm cls}=11$ elements as the machine predictions. Above each layer the height and width of its output are shown. The different layer depths reflect differences in their hyperparameters. Note that only one of the four repetitive layer patterns of the first block is shown for more clarity.	}
	\label{fig:arch}
\end{figure*} 


\subsection{Deep search}
\label{Deep_sec}

As the activation function, and in a tradeoff between performance and time-consumption, we use  the Concatenated Rectified Linear Unit or Concatenated  ReLU or CReLU that concatenates a ReLU function \citep{shang2016understanding}, in all but the output layer where Softmax is  used.
 The  ReLU function preserves only the positive part of the activation, $f(x)=\max(x,0)$. 
 CReLU, on the other hand,  concatenates a ReLU which selects only the positive part of the activation with a ReLU which selects only the negative part of the activation and has the notable feature of information preservation.
For the loss function, among the various  functions tried, we find the Huber-Loss to lead to best predictions,  
\begin{align}
L_{\delta}(y,f(x)) = \left\{ \begin{array}{cl}
\frac{1}{2} \left[y-f(x)\right]^2 & \text{for }|y-f(x)| \le \delta, \\
\delta \left(|y-f(x)|-\delta/2\right) & \text{otherwise.}
\end{array}\right.
\label{huber_loss}
\end{align}
For the learning rate or LR, we start with $0.001$ and gradually decrease it to $5\times 10^{-5}$ as the cost value settles down during the training process.
To deal with the destructive impact of 
 instrumental noise on the network accuracy, we 
can also use Transfer Learning methods \citep{pan2009survey}, where the learning process is transferred from noise-free  to noisy data. 
In these methods the network is first trained on noise-free datasets. 
Then some of the pre-trained layers are deactivated while the rest are allowed to be trained with the noisy data. 

The network coding of this work is done in the python environment, and heavily uses the TensorFlow \citep{abadi2016tensorflow} python library.
Details of the network were explained before. The final model has around $90,000$ free parameters describing the architecture and  the hyperparameters. 
The training process with the goal to find the best fit of all these parameters is done during about 4500 iterations where in each iteration the network sees around 100 images. So the total amount of data fed to the network is about 450000 patches, or close to $3\times10^{10}$ pixels\footnote{This training process takes about four hours to run on 12GB  Tesla K80 GPU.}.

The final architecture for the classifier used in this work is described in Table~\ref{Table:arch}. 
Figure~\ref{fig:arch} visualizes the network and how it reduces the dimension through 16 convolution layers of the input data (a $256\times 256$ image) into an array with $11$ components corresponding to the $11$ classes of $G\mu$.
The first two layers of this network are repeated four and six times respectively and they both have convolution and BN layers. The second block has  pooling and strides as well, and therefor reduces the input dimension. These convolution blocks are followed by a flatten layer to transform the 2D image into a 1D array which is fed into the fully connected layer (Dense). Also, $50\%$ of nodes are also dropped in dropout layers to avoid overfitting. The output of the very last Dense layer, with a Softmax activation, is a 1D array with 11 elements corresponding to the network predictions for the $G\mu$ probability of the input maps.

\begin{figure}
	\includegraphics[width=\linewidth]{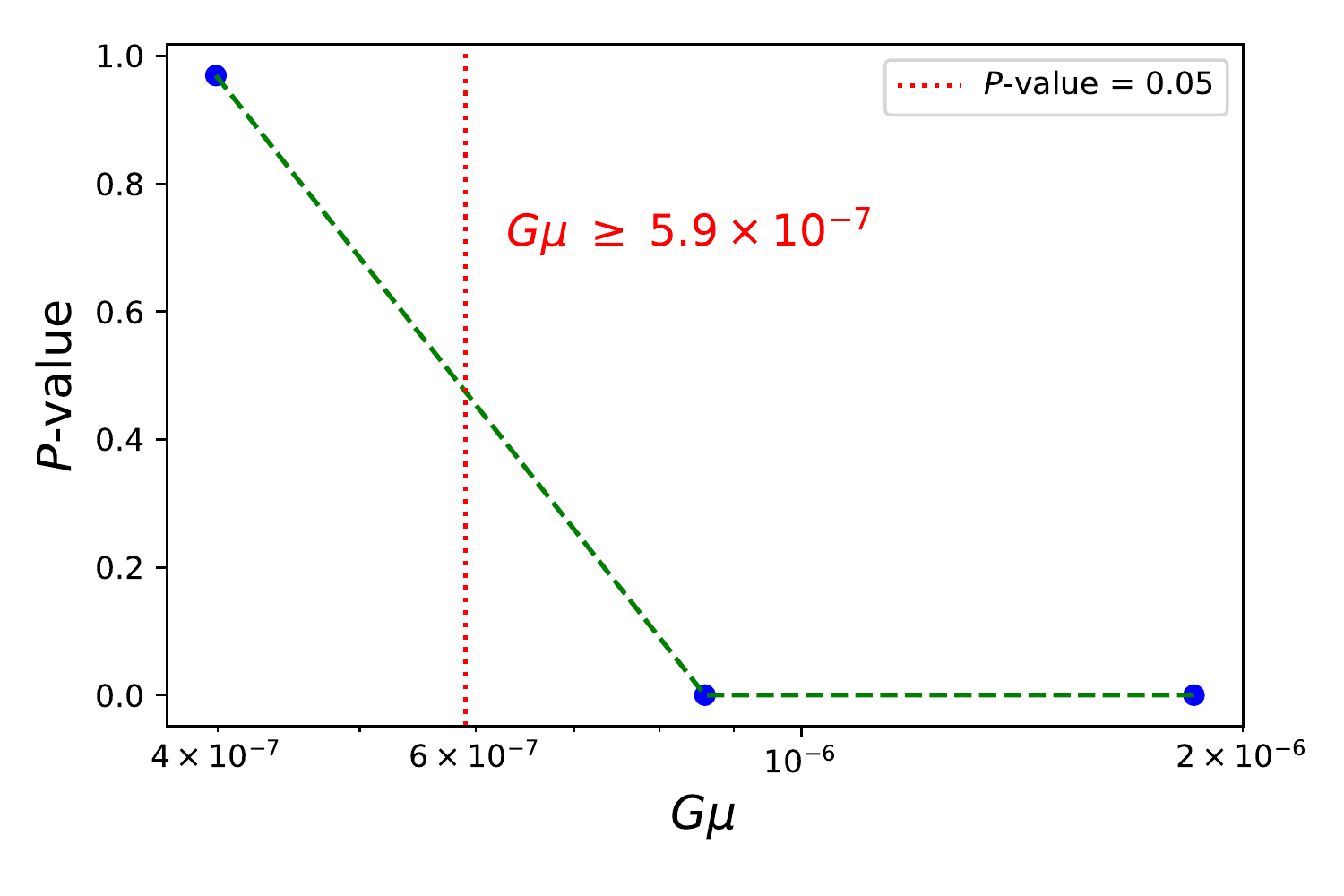}
	\caption{Similar to Figure~\ref{fig:pvallgbm} measured by the CNN on E2E simulations.}
	\label{fig:cnnpval}
\end{figure}

\begin{figure}
	\centering
	\includegraphics[width=0.47\textwidth ,height=0.35\textheight]{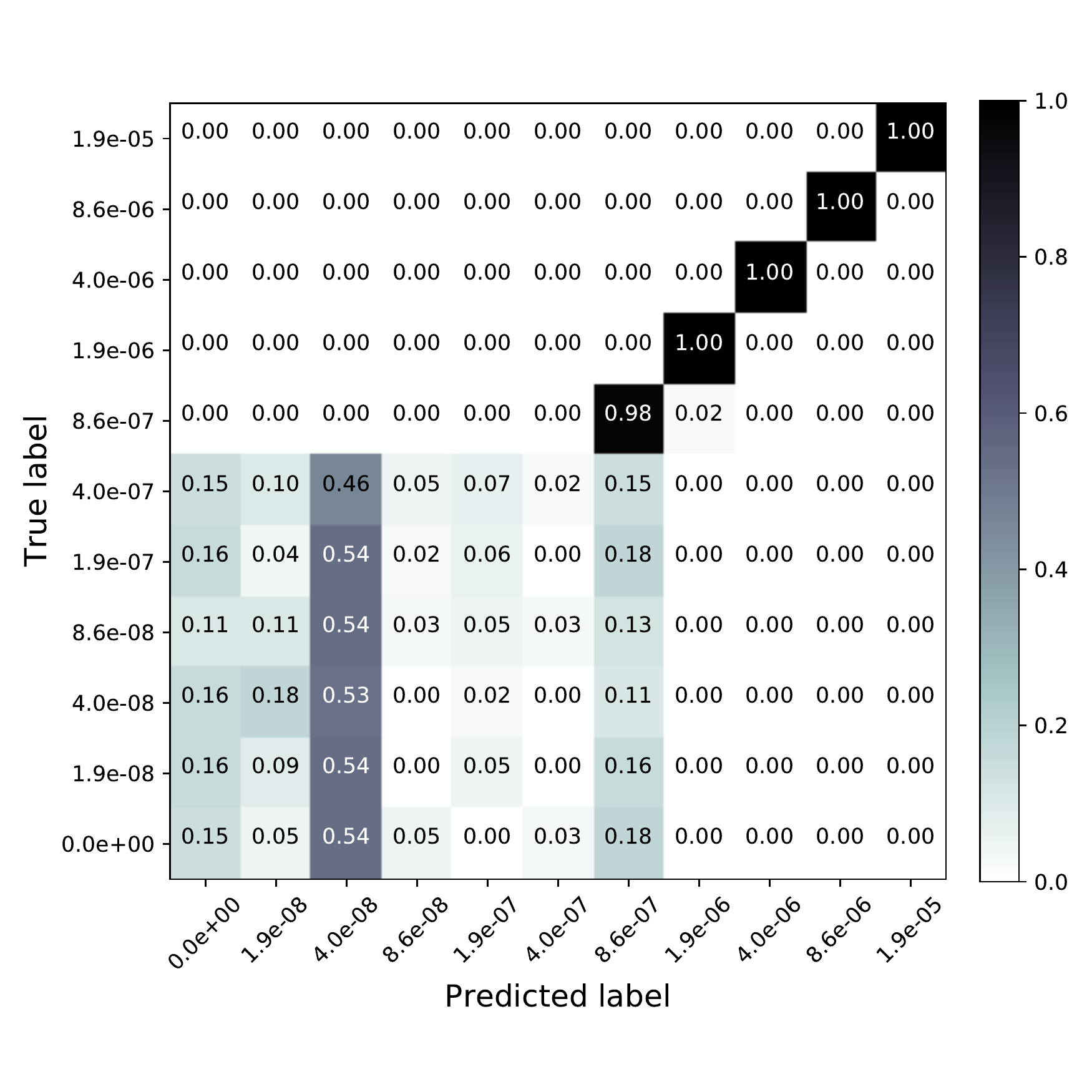}
	\caption{Similar to Figure~\ref{fig:conflgbm} measured by the CNN on E2E simulations.}
	\label{fig:cnnconf}
\end{figure}

To report the capability of the network in measuring the cosmic string signature, we use both the {\it P}-value statistics and the confusion matrix, similar to the case of educated search (Section~\ref{sec:edu}).
Figures~\ref{fig:cnnpval} and~\ref{fig:cnnconf} illustrate the performance of the final network  for unseen E2E simulations. From this confusion matrix, we find that the network can powerfully distinguish the class of $G\mu=8.6\times 10^{-7}$ and above from the null case and from each other, while it has not learned much about the $6\times6$ lower left block of the matrix. In this confused block, the majority of the predictions happen to be stacked in the third class. We therefore report $G\mu_{\rm min}=8.6\times10^{-7}$ based on the E2E simulations. 

\section{Results}\label{sec:res}
\begin{figure}
\centering
  \includegraphics[width=1.\columnwidth]{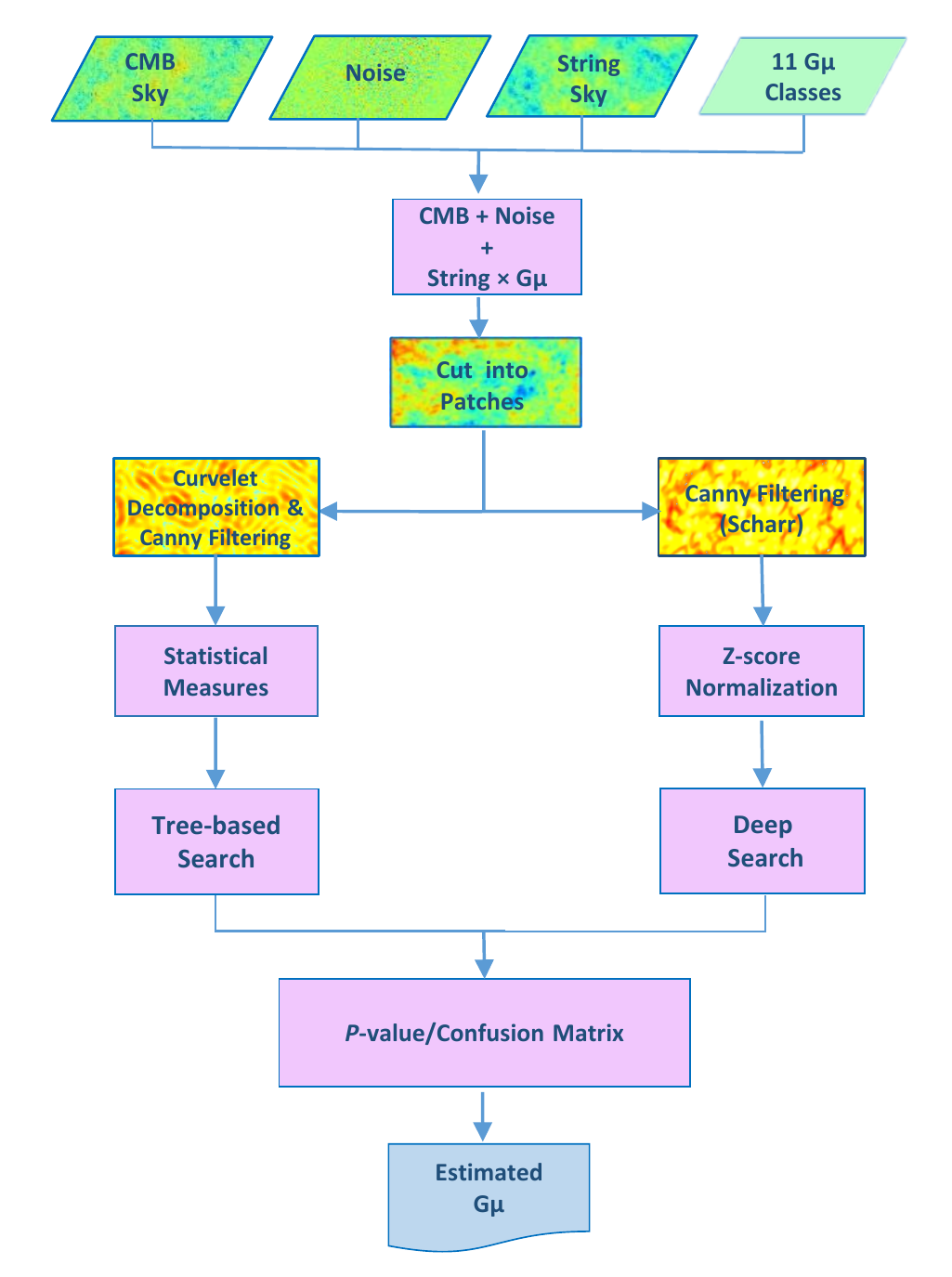}
  \caption{Overview of the data preparation and training pipelines of this work.}
  \label{fig:overview}
\end{figure}

The pipelines for educated and deep cosmic string search were described in section~\ref{sec:det} and developed based on simulations as introduced in Section~\ref{sec:sim}.
An overview of these algorithms is presented in Figure~\ref{fig:overview}.
The results of the analysis and training processes and the performance of the various paths in the pipeline for these different sets of simulations are presented in Table~\ref{table:res}. The middle and last columns  are based on the confusion matrix and  {\it P}-value measurements respectively. 
We see that the detectability limit from the confusion matrix is more conservative than the {\it P}-value. 
As was discussed in Section~\ref{sec:MLmethod}, while a direct comparison between the distribution of predictions for different classes can distinguish the class with  lower $G\mu$ imprints from the null case, the confusion matrix implies that the predictions for this low $G\mu$
region are not fully reliable. 
We therefore conservatively use the bound from the confusion matrix as the main limit.
With this measure the deep search has often a better performance for cases with relatively low noise level. 

We apply the pipelines developed for the E2E {\it Planck} simulations to the \texttt{ SMICA} temperature map.
We divide the full (except for the mask) sky {\it Planck} map into smaller patches, as was done for the simulations, and 
compare the histogram of predictions for both pipelines with the histograms of the individual $G\mu$ classes from simulations through calculating the {\it P}-value for each case. The measured {\it P}-values for all cases are below  $0.0027$ which corresponds to the $3\sigma$ frequentist level.
We therefore conclude that $G\mu$ is below the minimum detectability limit of the network with $3\sigma$ significance.  The $3\sigma$ upper bound on the  $G\mu$ as estimated from  {\it Planck} sky is $G\mu \lesssim 8.6 \times 10^{-7}$, for both LightGBM and deep search methods.

We also explored binary classifications with two classes of $G\mu=0$ as  class zero, and $G\mu \gtrsim G\mu_{\rm min}$ as class one, to see whether the upper bounds can be improved. However, we found no significant improvement over the results  of Table~\ref{table:res} for the E2E case and therefore the $G\mu \lesssim 8.6 \times 10^{-7}$ bound is our final result.
\begin{table}
	\captionof{table}{The minimum detectable $G\mu_{\rm min}(\times 10^{-7})$ from LightGBM and CNN pipelines based on the confusion matrix and {\it P}-value measurements for different sets of simulations.}
	\label{table:res}
	\begin{tabularx}{0.5\textwidth}{>{\raggedleft}X|>{\centering}XX|>{\centering}XX|}
		\cline{2-5}
		& \multicolumn{2}{c|}{ Confusion Matrix} & \multicolumn{2}{c|}{ {\it P}-value}  \\ \hline
		\multicolumn{1}{|c|}{Experiment}            & LGBM  & CNN  & LGBM & CNN        \\
		\hline
		
		\multicolumn{1}{|c|}{CMB-S4-like(II)}       &  4.0 & 1.9  & 0.5  & 0.8   \\
		\multicolumn{1}{|c|}{CMB-S4-like(I)}        &  4.0  & 1.9  & 0.8  & 1.2   \\
		\multicolumn{1}{|c|}{noise-free FFP10}      &  4.0  & 0.4  & 0.3  & 0.3  \\
		\multicolumn{1}{|c|}{FFP10}                 &  8.6  & 8.6  & 1.7  & 3.6   \\
		\multicolumn{1}{|c|}{noise-free E2E}        &  8.6  & 8.6  &  2.1 & 3.8   \\
        \multicolumn{1}{|c|}{ E2E}                  &  8.6  & 8.6  & 2.7  & 5.9   \\       \hline
	\end{tabularx}
\end{table}
 \section{Summary and discussion}\label{sec:sum}
In this work we developed two parallel tree-based and deep pipelines with the goal to estimate or put tight upper bounds on the level of cosmic string contribution
to the observed CMB anisotropies by {\it Planck}18 data. We also made forecasts for the $G\mu$ detectability by next generation CMB experiments.  
Table~\ref{table:res} summarizes the performance of the pipelines  for these various observational scenarios using two different criteria, here labeled by {\it P}-value and confusion matrix. 
The limits from the confusion matrix are more conservative by making sure that the predictions of the trained models are not driven  by confused decisions. 
Overall, the detectability limits based on the {\it P}-value and confusion matrix of Table~\ref{table:res} are respectively comparable with and slightly higher than the previous machine-based forecasts of \cite{vaf18}. In particular, both the LightGBM and CNN pipelines put tighter bounds on string tension in the case of noisy {\it Planck} simulations. 

We used the machines trained with E2E {\it Planck} simulations to search for the CS imprints on the observed \texttt{SMICA} map. Both LightGBM and CNN models found no observable trace  and yielded $G\mu \lesssim 8.6 \times 10^{-7}$  with $3\sigma$ confidence level.
It should be noted that these bounds correspond to the discrete class labels in the confusion matrix and are therefore conservative. The true upper bound could be in between the current bound and the class label right below it, which needs to be  investigated with finer grids in the vicinity of the current upper bound. 
Further improvement could also be achieved through improving the string simulations. In our analysis it would require the string maps to go  through the very same pipeline as the primordial CMB maps, such as E2E and FFP10. 
Here, instead, we have  used  an effective Gaussian beam.  This could have slightly deteriorated the machines power in learning about the true string signal.
Training on more string simulations is also expected to help the machines in distinguishing the string imprints from background and instrumental noise. Accurate simulations of CS-induced anisotropies are quite expensive, which motivates trying producing simulations from the Generative  Adversarial Network methods. 

In parallel to the geometrical and statistical features and image processing tools applied in this work, one could 
 explore topological measures and invariants through, e.g., investigating the $k$-homology classes
\citep{matsubara2003statistics,adler1981geometry,adler2011topological,adler2010persistent,kozlov2007combinatorial,bubenik2015statistical}.
The topological measures could then be included in the feature vector to improve the learning process and the precision of the predictions. The full exploration of these steps are left to future work.


\section*{Acknowledgment}
The authors are grateful to  C. Ringeval, F. R. Bouchet, and the {\it Planck} collaboration as they generously provided the simulations of the cosmic string maps used in this work.
The numerical computations were carried out on the Baobab cluster at University of Geneva and the Google collaboratory facilities.

\bibliographystyle{mnras}
\bibliography{bib} 

\begin{thebibliography}{}
\makeatletter
\relax
\def\mn@urlcharsother{\let\do\@makeother \do\$\do\&\do\#\do\^\do\_\do\%\do\~}
\def\mn@doi{\begingroup\mn@urlcharsother \@ifnextchar [ {\mn@doi@}
  {\mn@doi@[]}}
\def\mn@doi@[#1]#2{\def\@tempa{#1}\ifx\@tempa\@empty \href
  {http://dx.doi.org/#2} {doi:#2}\else \href {http://dx.doi.org/#2} {#1}\fi
  \endgroup}
\def\mn@eprint#1#2{\mn@eprint@#1:#2::\@nil}
\def\mn@eprint@arXiv#1{\href {http://arxiv.org/abs/#1} {{\tt arXiv:#1}}}
\def\mn@eprint@dblp#1{\href {http://dblp.uni-trier.de/rec/bibtex/#1.xml}
  {dblp:#1}}
\def\mn@eprint@#1:#2:#3:#4\@nil{\def\@tempa {#1}\def\@tempb {#2}\def\@tempc
  {#3}\ifx \@tempc \@empty \let \@tempc \@tempb \let \@tempb \@tempa \fi \ifx
  \@tempb \@empty \def\@tempb {arXiv}\fi \@ifundefined
  {mn@eprint@\@tempb}{\@tempb:\@tempc}{\expandafter \expandafter \csname
  mn@eprint@\@tempb\endcsname \expandafter{\@tempc}}}

\bibitem[\protect\citeauthoryear{Abadi et~al.,}{Abadi
  et~al.}{2016}]{abadi2016tensorflow}
Abadi M.,  et~al., 2016, in 12th $\{$USENIX$\}$ symposium on operating systems
  design and implementation ($\{$OSDI$\}$ 16). pp 265--283

\bibitem[\protect\citeauthoryear{Ade et~al.,}{Ade et~al.}{2014}]{ade2014planck}
Ade P.~A.,  et~al., 2014, Astronomy \& Astrophysics, 571, A25

\bibitem[\protect\citeauthoryear{Ade et~al.,}{Ade et~al.}{2016}]{ade2016planck}
Ade P.~A.,  et~al., 2016, Astronomy \& Astrophysics, 594, A13

\bibitem[\protect\citeauthoryear{Adler}{Adler}{1981}]{adler1981geometry}
Adler R.~J.,  1981, SIAM, Philadelphia

\bibitem[\protect\citeauthoryear{Adler \& Taylor}{Adler \&
  Taylor}{2011}]{adler2011topological}
Adler R.,  Taylor J.~E.,  2011, Topological Complexity of Smooth Random
  Functions: {\'E}cole D'{\'E}t{\'e} de Probabilit{\'e}s de Saint-Flour
  XXXIX-2009.
Springer Science \& Business Media

\bibitem[\protect\citeauthoryear{Adler, Bobrowski, Borman, Subag, Weinberger
  et~al.}{Adler et~al.}{2010}]{adler2010persistent}
Adler R.~J.,  Bobrowski O.,  Borman M.~S.,  Subag E.,  Weinberger S.,   et~al.,
  2010, in , Borrowing strength: theory powering applications--a Festschrift
  for Lawrence D. Brown.
Institute of Mathematical Statistics, pp 124--143

\bibitem[\protect\citeauthoryear{Akrami et~al.,}{Akrami
  et~al.}{2019}]{akrami2019planck}
Akrami Y.,  et~al., 2019, arXiv preprint arXiv:1906.02552

\bibitem[\protect\citeauthoryear{Allen, Knox, Shellard, Caldwell, Dodelson  \&
  Stebbins}{Allen et~al.}{1997}]{allen1997cmb}
Allen B.,  Knox L.,  Shellard E.,  Caldwell R.,  Dodelson S.,   Stebbins A.,
  1997, Phys. Rev. Lett., 79, 2624

\bibitem[\protect\citeauthoryear{Aloysius \& Geetha}{Aloysius \&
  Geetha}{2017}]{aloysius2017review}
Aloysius N.,  Geetha M.,  2017, in 2017 International Conference on
  Communication and Signal Processing (ICCSP). pp 0588--0592

\bibitem[\protect\citeauthoryear{Arzoumanian et~al.,}{Arzoumanian
  et~al.}{2018}]{arz18}
Arzoumanian Z.,  et~al., 2018, \mn@doi [The Astrophysical Journal]
  {10.3847/1538-4357/aabd3b}, 859, 47

\bibitem[\protect\citeauthoryear{Auclair et~al.,}{Auclair et~al.}{2020}]{auc19}
Auclair P.,  et~al., 2020, \mn@doi [Journal of Cosmology and Astroparticle
  Physics] {10.1088/1475-7516/2020/04/034}, 2020, 034?034

\bibitem[\protect\citeauthoryear{Battye \& Moss}{Battye \& Moss}{2010}]{bat10}
Battye R.,  Moss A.,  2010, \mn@doi [Phys. Rev.] {10.1103/PhysRevD.82.023521},
  D82, 023521

\bibitem[\protect\citeauthoryear{Bennett \& Bouchet}{Bennett \&
  Bouchet}{1990}]{bennett1990high}
Bennett D.~P.,  Bouchet F.~R.,  1990, Physical Review D, 41, 2408

\bibitem[\protect\citeauthoryear{Bevis, Hindmarsh, Kunz  \& Urrestilla}{Bevis
  et~al.}{2008}]{Bevis:2007gh}
Bevis N.,  Hindmarsh M.,  Kunz M.,   Urrestilla J.,  2008, \mn@doi [Phys. Rev.
  Lett.] {10.1103/PhysRevLett.100.021301}, 100, 021301

\bibitem[\protect\citeauthoryear{Bevis, Hindmarsh, Kunz  \& Urrestilla}{Bevis
  et~al.}{2010}]{Bevis:2010gj}
Bevis N.,  Hindmarsh M.,  Kunz M.,   Urrestilla J.,  2010, \mn@doi [Phys. Rev.]
  {10.1103/PhysRevD.82.065004}, D82, 065004

\bibitem[\protect\citeauthoryear{Blanco-Pillado \& Olum}{Blanco-Pillado \&
  Olum}{2017}]{bla17a}
Blanco-Pillado J.~J.,  Olum K.~D.,  2017, Physical Review D, 96, 104046

\bibitem[\protect\citeauthoryear{Blanco-Pillado, Olum  \&
  Siemens}{Blanco-Pillado et~al.}{2017}]{bla17b}
Blanco-Pillado J.~J.,  Olum K.~D.,   Siemens X.,  2017, arXiv preprint
  arXiv:1709.02434

\bibitem[\protect\citeauthoryear{Blanco-Pillado, Olum  \&
  Siemens}{Blanco-Pillado et~al.}{2018}]{bla18}
Blanco-Pillado J.~J.,  Olum K.~D.,   Siemens X.,  2018, \mn@doi [Phys. Lett. B]
  {10.1016/j.physletb.2018.01.050}, 778, 392

\bibitem[\protect\citeauthoryear{Bouchet, Bennett  \& Stebbins}{Bouchet
  et~al.}{1988}]{bouchet1988microwave}
Bouchet F.~R.,  Bennett D.~P.,   Stebbins A.,  1988, Nature, 335, 410

\bibitem[\protect\citeauthoryear{Brandenberger, Danos, Hernandez  \&
  Holder}{Brandenberger et~al.}{2010}]{bra10}
Brandenberger R.~H.,  Danos R.~J.,  Hernandez O.~F.,   Holder G.~P.,  2010,
  \mn@doi [JCAP] {10.1088/1475-7516/2010/12/028}, 1012, 028

\bibitem[\protect\citeauthoryear{Breiman}{Breiman}{2001}]{bre01}
Breiman L.,  2001, Machine learning, 45, 5

\bibitem[\protect\citeauthoryear{Bubenik}{Bubenik}{2015}]{bubenik2015statistical}
Bubenik P.,  2015, The Journal of Machine Learning Research, 16, 77

\bibitem[\protect\citeauthoryear{Candes \& Donoho}{Candes \&
  Donoho}{2000}]{can00}
Candes E.~J.,  Donoho D.~L.,  2000, Technical report, Curvelets: A surprisingly
  effective nonadaptive representation for objects with edges.
DTIC Document

\bibitem[\protect\citeauthoryear{Cand{\`e}s \& Donoho}{Cand{\`e}s \&
  Donoho}{2001}]{can01}
Cand{\`e}s E.~J.,  Donoho D.~L.,  2001, Journal of Approximation Theory, 113,
  59

\bibitem[\protect\citeauthoryear{Cand{\`e}s \& Guo}{Cand{\`e}s \&
  Guo}{2002}]{can02}
Cand{\`e}s E.~J.,  Guo F.,  2002, Signal Processing, 82, 1519

\bibitem[\protect\citeauthoryear{Candes, Demanet, Donoho  \& Ying}{Candes
  et~al.}{2006}]{can06}
Candes E.,  Demanet L.,  Donoho D.,   Ying L.,  2006, Multiscale Modeling \&
  Simulation, 5, 861

\bibitem[\protect\citeauthoryear{Canny}{Canny}{1986}]{can86}
Canny J.,  1986, IEEE Transactions on pattern analysis and machine
  intelligence, pp 679--698

\bibitem[\protect\citeauthoryear{Charnock, Avgoustidis, Copeland  \&
  Moss}{Charnock et~al.}{2016}]{cha16}
Charnock T.,  Avgoustidis A.,  Copeland E.~J.,   Moss A.,  2016, \mn@doi [Phys.
  Rev.] {10.1103/PhysRevD.93.123503}, D93, 123503

\bibitem[\protect\citeauthoryear{Chen \& Guestrin}{Chen \&
  Guestrin}{2016}]{che16}
Chen T.,  Guestrin C.,  2016, in Proceedings of the 22nd acm sigkdd
  international conference on knowledge discovery and data mining. pp 785--794

\bibitem[\protect\citeauthoryear{Ciregan, Meier  \& Schmidhuber}{Ciregan
  et~al.}{2012}]{cir12}
Ciregan D.,  Meier U.,   Schmidhuber J.,  2012, in 2012 IEEE conference on
  computer vision and pattern recognition. pp 3642--3649

\bibitem[\protect\citeauthoryear{Ciuca \& Hern{\'a}ndez}{Ciuca \&
  Hern{\'a}ndez}{2017}]{ciu17}
Ciuca R.,  Hern{\'a}ndez O.~F.,  2017, Journal of Cosmology and Astroparticle
  Physics, 2017, 028

\bibitem[\protect\citeauthoryear{Ciuca \& Hern{\'a}ndez}{Ciuca \&
  Hern{\'a}ndez}{2020}]{ciu20}
Ciuca R.,  Hern{\'a}ndez O.~F.,  2020, Monthly Notices of the Royal
  Astronomical Society, 492, 1329

\bibitem[\protect\citeauthoryear{Copeland, Liddle, Lyth, Stewart  \&
  Wands}{Copeland et~al.}{1994}]{Copeland:1994vg}
Copeland E.~J.,  Liddle A.~R.,  Lyth D.~H.,  Stewart E.~D.,   Wands D.,  1994,
  \mn@doi [Phys. Rev.] {10.1103/PhysRevD.49.6410}, D49, 6410

\bibitem[\protect\citeauthoryear{Copeland, Myers  \& Polchinski}{Copeland
  et~al.}{2004}]{Copeland:2003bj}
Copeland E.~J.,  Myers R.~C.,   Polchinski J.,  2004, \mn@doi [JHEP]
  {10.1088/1126-6708/2004/06/013}, 06, 013

\bibitem[\protect\citeauthoryear{Cunha, Harnois-Deraps, Brandenberger, Amara
  \& Refregier}{Cunha et~al.}{2018}]{cun18}
Cunha D.,  Harnois-Deraps J.,  Brandenberger R.,  Amara A.,   Refregier A.,
  2018, Dark Matter Distribution Induced by a Cosmic String Wake in the
  Nonlinear Regime (\mn@eprint {arXiv} {1804.00083})

\bibitem[\protect\citeauthoryear{Damour \& Vilenkin}{Damour \&
  Vilenkin}{2005}]{dam05}
Damour T.,  Vilenkin A.,  2005, \mn@doi [Phys. Rev.]
  {10.1103/PhysRevD.71.063510}, D71, 063510

\bibitem[\protect\citeauthoryear{Depies}{Depies}{2009}]{Depies:2009im}
Depies M.~R.,  2009, arXiv preprint arXiv:0908.3680

\bibitem[\protect\citeauthoryear{Donoho \& Duncan}{Donoho \&
  Duncan}{2000}]{don00}
Donoho D.~L.,  Duncan M.~R.,  2000, in AeroSense 2000. pp 12--30

\bibitem[\protect\citeauthoryear{Ducout, Bouchet, Colombi, Pogosyan  \&
  Prunet}{Ducout et~al.}{2013}]{duc12}
Ducout A.,  Bouchet F.,  Colombi S.,  Pogosyan D.,   Prunet S.,  2013, \mn@doi
  [Mon. Not. Roy. Astron. Soc.] {10.1093/mnras/sts483}, 429, 2104

\bibitem[\protect\citeauthoryear{Dvali \& Vilenkin}{Dvali \&
  Vilenkin}{2004}]{Dvali:2003zj}
Dvali G.,  Vilenkin A.,  2004, \mn@doi [JCAP] {10.1088/1475-7516/2004/03/010},
  0403, 010

\bibitem[\protect\citeauthoryear{Fernandez, Bird  \& Cui}{Fernandez
  et~al.}{2020}]{fer20}
Fernandez M.,  Bird S.,   Cui Y.,  2020, arXiv preprint arXiv:2004.13752

\bibitem[\protect\citeauthoryear{Goodfellow, Bengio  \& Courville}{Goodfellow
  et~al.}{2016}]{goodfellow2016deep}
Goodfellow I.,  Bengio Y.,   Courville A.,  2016, Deep learning.
MIT press

\bibitem[\protect\citeauthoryear{{Gott}~III}{{Gott}~III}{1985}]{Gott:1985}
{Gott}~III J.~R.,  1985, \mn@doi [\apj] {10.1086/162808}, \href
  {http://adsabs.harvard.edu/abs/1985ApJ...288..422G} {288, 422}

\bibitem[\protect\citeauthoryear{Gu et~al.,}{Gu et~al.}{2018}]{gu2018recent}
Gu J.,  et~al., 2018, Pattern Recognition, 77, 354

\bibitem[\protect\citeauthoryear{Hassoun et~al.}{Hassoun
  et~al.}{1995}]{hassoun1995fundamentals}
Hassoun M.~H.,  et~al., 1995, Fundamentals of artificial neural networks.
MIT press

\bibitem[\protect\citeauthoryear{Henry~Tye}{Henry~Tye}{2008}]{HenryTye:2006uv}
Henry~Tye S.~H.,  2008, Lect. Notes Phys., 737, 949

\bibitem[\protect\citeauthoryear{Hergt, Amara, Brandenberger, Kacprzak  \&
  R{\'e}fr{\'e}gier}{Hergt et~al.}{2017}]{her16}
Hergt L.,  Amara A.,  Brandenberger R.,  Kacprzak T.,   R{\'e}fr{\'e}gier A.,
  2017, Journal of Cosmology and Astroparticle Physics, 2017, 004

\bibitem[\protect\citeauthoryear{Hernandez \& Brandenberger}{Hernandez \&
  Brandenberger}{2012}]{her12}
Hernandez O.~F.,  Brandenberger R.~H.,  2012, \mn@doi [JCAP]
  {10.1088/1475-7516/2012/07/032}, 1207, 032

\bibitem[\protect\citeauthoryear{Hindmarsh}{Hindmarsh}{1994}]{Hindmarsh:1993pu}
Hindmarsh M.,  1994, \mn@doi [Astrophys. J.] {10.1086/174505}, 431, 534

\bibitem[\protect\citeauthoryear{Hindmarsh \& Kibble}{Hindmarsh \&
  Kibble}{1995}]{Hindmarsh:1994re}
Hindmarsh M.~B.,  Kibble T. W.~B.,  1995, \mn@doi [Rept. Prog. Phys.]
  {10.1088/0034-4885/58/5/001}, 58, 477

\bibitem[\protect\citeauthoryear{Hindmarsh, Ringeval  \& Suyama}{Hindmarsh
  et~al.}{2009}]{hin09}
Hindmarsh M.,  Ringeval C.,   Suyama T.,  2009, \mn@doi [Phys. Rev.]
  {10.1103/PhysRevD.80.083501}, D80, 083501

\bibitem[\protect\citeauthoryear{Hindmarsh, Ringeval  \& Suyama}{Hindmarsh
  et~al.}{2010}]{hin10}
Hindmarsh M.,  Ringeval C.,   Suyama T.,  2010, \mn@doi [Phys. Rev.]
  {10.1103/PhysRevD.81.063505}, D81, 063505

\bibitem[\protect\citeauthoryear{Hinton, Srivastava, Krizhevsky, Sutskever  \&
  Salakhutdinov}{Hinton et~al.}{2012}]{hinton2012improving}
Hinton G.~E.,  Srivastava N.,  Krizhevsky A.,  Sutskever I.,   Salakhutdinov
  R.~R.,  2012, arXiv preprint arXiv:1207.0580

\bibitem[\protect\citeauthoryear{Imtiaz, Shi  \& Cai}{Imtiaz
  et~al.}{2020}]{imt20}
Imtiaz B.,  Shi R.,   Cai Y.-F.,  2020, \mn@doi [The European Physical Journal
  C] {10.1140/epjc/s10052-020-8064-x}, 80

\bibitem[\protect\citeauthoryear{Ioffe \& Szegedy}{Ioffe \&
  Szegedy}{2015}]{iof15}
Ioffe S.,  Szegedy C.,  2015, arXiv preprint arXiv:1502.03167

\bibitem[\protect\citeauthoryear{Jain, Mao  \& Mohiuddin}{Jain
  et~al.}{1996}]{jain1996artificial}
Jain A.~K.,  Mao J.,   Mohiuddin K.~M.,  1996, Computer, 29, 31

\bibitem[\protect\citeauthoryear{Jenet et~al.,}{Jenet et~al.}{2006}]{jen06}
Jenet F.~A.,  et~al., 2006, \mn@doi [Astrophys. J.] {10.1086/508702}, 653, 1571

\bibitem[\protect\citeauthoryear{Kaiser \& Stebbins}{Kaiser \&
  Stebbins}{1984}]{Kaiser:1984iv}
Kaiser N.,  Stebbins A.,  1984, \mn@doi [Nature] {10.1038/310391a0}, 310, 391

\bibitem[\protect\citeauthoryear{Ke, Meng, Finley, Wang, Chen, Ma, Ye  \&
  Liu}{Ke et~al.}{2017a}]{lig17}
Ke G.,  Meng Q.,  Finley T.,  Wang T.,  Chen W.,  Ma W.,  Ye Q.,   Liu T.-Y.,
  2017a, in Advances in neural information processing systems. pp 3146--3154

\bibitem[\protect\citeauthoryear{Ke, Meng, Finley, Wang, Chen, Ma, Ye  \&
  Liu}{Ke et~al.}{2017b}]{ke17}
Ke G.,  Meng Q.,  Finley T.,  Wang T.,  Chen W.,  Ma W.,  Ye Q.,   Liu T.-Y.,
  2017b, in Advances in neural information processing systems. pp 3146--3154

\bibitem[\protect\citeauthoryear{Keller, Gray  \& Givens}{Keller
  et~al.}{1985}]{kel85}
Keller J.~M.,  Gray M.~R.,   Givens J.~A.,  1985, IEEE transactions on systems,
  man, and cybernetics, pp 580--585

\bibitem[\protect\citeauthoryear{Kibble}{Kibble}{1976}]{Kibble:1976sj}
Kibble T. W.~B.,  1976, \mn@doi [J. Phys.] {10.1088/0305-4470/9/8/029}, A9,
  1387

\bibitem[\protect\citeauthoryear{Kibble}{Kibble}{2004}]{Kibble:2004hq}
Kibble T. W.~B.,  2004, in {COSLAB 2004 Ambleside, Cumbria, United Kingdom,
  September 10-17, 2004}.  (\mn@eprint {arXiv} {astro-ph/0410073})

\bibitem[\protect\citeauthoryear{Kozlov}{Kozlov}{2007}]{kozlov2007combinatorial}
Kozlov D.,  2007, Combinatorial algebraic topology.
 Vol. 21, Springer Science \& Business Media

\bibitem[\protect\citeauthoryear{Krizhevsky, Sutskever  \& Hinton}{Krizhevsky
  et~al.}{2012}]{krizhevsky2012imagenet}
Krizhevsky A.,  Sutskever I.,   Hinton G.~E.,  2012, in Advances in neural
  information processing systems. pp 1097--1105

\bibitem[\protect\citeauthoryear{Kuroyanagi, Miyamoto, Sekiguchi, Takahashi  \&
  Silk}{Kuroyanagi et~al.}{2013}]{kur12}
Kuroyanagi S.,  Miyamoto K.,  Sekiguchi T.,  Takahashi K.,   Silk J.,  2013,
  \mn@doi [Phys. Rev.] {10.1103/PhysRevD.87.069903,
  10.1103/PhysRevD.87.023522}, D87, 023522

\bibitem[\protect\citeauthoryear{Laliberte \& Brandenberger}{Laliberte \&
  Brandenberger}{2020}]{lal19}
Laliberte S.,  Brandenberger R.,  2020, \mn@doi [Physical Review D]
  {10.1103/physrevd.101.023528}, 101

\bibitem[\protect\citeauthoryear{Lazanu \& Shellard}{Lazanu \&
  Shellard}{2015}]{lazanu2015constraints}
Lazanu A.,  Shellard P.,  2015, Journal of Cosmology and Astroparticle Physics,
  2015, 024

\bibitem[\protect\citeauthoryear{Lentati et~al.,}{Lentati et~al.}{2015}]{len15}
Lentati L.,  et~al., 2015, \mn@doi [Monthly Notices of the Royal Astronomical
  Society] {10.1093/mnras/stv1538}, 453, 2577?2599

\bibitem[\protect\citeauthoryear{Majumdar \& Christine-Davis}{Majumdar \&
  Christine-Davis}{2002}]{Majumdar:2002hy}
Majumdar M.,  Christine-Davis A.,  2002, \mn@doi [JHEP]
  {10.1088/1126-6708/2002/03/056}, 03, 056

\bibitem[\protect\citeauthoryear{Matsubara}{Matsubara}{2003}]{matsubara2003statistics}
Matsubara T.,  2003, The Astrophysical Journal, 584, 1

\bibitem[\protect\citeauthoryear{Mittal}{Mittal}{2020}]{mit20}
Mittal S.,  2020, Neural computing and applications, pp 1--31

\bibitem[\protect\citeauthoryear{Movahed \& Khosravi}{Movahed \&
  Khosravi}{2011}]{mov10}
Movahed M.~S.,  Khosravi S.,  2011, \mn@doi [JCAP]
  {10.1088/1475-7516/2011/03/012}, 1103, 012

\bibitem[\protect\citeauthoryear{Movahed, Javanmardi  \& Sheth}{Movahed
  et~al.}{2012}]{mov12}
Movahed M.~S.,  Javanmardi B.,   Sheth R.~K.,  2012, \mn@doi [Mon. Not. Roy.
  Astron. Soc.] {10.1093/mnras/stt1284}, 434, 3597

\bibitem[\protect\citeauthoryear{Nwankpa, Ijomah, Gachagan  \&
  Marshall}{Nwankpa et~al.}{2018}]{nwankpa2018activation}
Nwankpa C.,  Ijomah W.,  Gachagan A.,   Marshall S.,  2018, arXiv preprint
  arXiv:1811.03378

\bibitem[\protect\citeauthoryear{Pagano \& Brandenberger}{Pagano \&
  Brandenberger}{2012}]{pag12}
Pagano M.,  Brandenberger R.,  2012, \mn@doi [Journal of Cosmology and
  Astroparticle Physics] {10.1088/1475-7516/2012/05/014}, 2012, 014?014

\bibitem[\protect\citeauthoryear{Pan \& Yang}{Pan \&
  Yang}{2009}]{pan2009survey}
Pan S.~J.,  Yang Q.,  2009, IEEE Transactions on knowledge and data
  engineering, 22, 1345

\bibitem[\protect\citeauthoryear{Pen, Seljak  \& Turok}{Pen
  et~al.}{1997}]{pen1997power}
Pen U.-L.,  Seljak U.,   Turok N.,  1997, Physical Review Letters, 79, 1611

\bibitem[\protect\citeauthoryear{Pogosian, Tye, Wasserman  \& Wyman}{Pogosian
  et~al.}{2003}]{Pogosian:2003mz}
Pogosian L.,  Tye S. H.~H.,  Wasserman I.,   Wyman M.,  2003, \mn@doi [Phys.
  Rev.] {10.1103/PhysRevD.68.023506, 10.1103/PhysRevD.73.089904}, D68, 023506

\bibitem[\protect\citeauthoryear{Pshirkov \& Tuntsov}{Pshirkov \&
  Tuntsov}{2010}]{psh09}
Pshirkov M.~S.,  Tuntsov A.~V.,  2010, \mn@doi [Phys. Rev.]
  {10.1103/PhysRevD.81.083519}, D81, 083519

\bibitem[\protect\citeauthoryear{Quinlan}{Quinlan}{1986}]{qui86}
Quinlan J.~R.,  1986, Machine learning, 1, 81

\bibitem[\protect\citeauthoryear{Regan \& Hindmarsh}{Regan \&
  Hindmarsh}{2015}]{reg15}
Regan D.,  Hindmarsh M.,  2015, \mn@doi [JCAP] {10.1088/1475-7516/2015/10/030},
  1510, 030

\bibitem[\protect\citeauthoryear{Ringeval}{Ringeval}{2010}]{rin10}
Ringeval C.,  2010, \mn@doi [Adv. Astron.] {10.1155/2010/380507}, 2010, 380507

\bibitem[\protect\citeauthoryear{Ringeval \& Bouchet}{Ringeval \&
  Bouchet}{2012}]{ringeval2012all}
Ringeval C.,  Bouchet F.~R.,  2012, Physical Review D, 86, 023513

\bibitem[\protect\citeauthoryear{Ringeval \& Suyama}{Ringeval \&
  Suyama}{2017}]{rin17}
Ringeval C.,  Suyama T.,  2017, Journal of Cosmology and Astroparticle Physics,
  2017, 027

\bibitem[\protect\citeauthoryear{Ringeval, Sakellariadou  \& Bouchet}{Ringeval
  et~al.}{2007}]{ringeval2007cosmological}
Ringeval C.,  Sakellariadou M.,   Bouchet F.~R.,  2007, Journal of Cosmology
  and Astroparticle Physics, 2007, 023

\bibitem[\protect\citeauthoryear{Rish et~al.}{Rish et~al.}{2001}]{ris01}
Rish I.,  et~al., 2001, in IJCAI 2001 workshop on empirical methods in
  artificial intelligence. pp 41--46

\bibitem[\protect\citeauthoryear{Sakellariadou}{Sakellariadou}{1997}]{Sakellariadou:1997zt}
Sakellariadou M.,  1997, \mn@doi [Int. J. Theor. Phys.] {10.1007/BF02768939},
  36, 2503

\bibitem[\protect\citeauthoryear{Sakellariadou}{Sakellariadou}{2007}]{Sakellariadou:2006qs}
Sakellariadou M.,  2007, \mn@doi [Lect. Notes Phys.]
  {10.1007/3-540-70859-6_10}, 718, 247

\bibitem[\protect\citeauthoryear{Sarangi \& Tye}{Sarangi \&
  Tye}{2002}]{Sarangi:2002yt}
Sarangi S.,  Tye S. H.~H.,  2002, \mn@doi [Phys. Lett.]
  {10.1016/S0370-2693(02)01824-5}, B536, 185

\bibitem[\protect\citeauthoryear{Shang, Sohn, Almeida  \& Lee}{Shang
  et~al.}{2016}]{shang2016understanding}
Shang W.,  Sohn K.,  Almeida D.,   Lee H.,  2016, in international conference
  on machine learning. pp 2217--2225

\bibitem[\protect\citeauthoryear{Shellard}{Shellard}{1987}]{Shellard:1987bv}
Shellard E. P.~S.,  1987, \mn@doi [Nucl. Phys.] {10.1016/0550-3213(87)90290-2},
  B283, 624

\bibitem[\protect\citeauthoryear{Srivastava, Hinton, Krizhevsky, Sutskever  \&
  Salakhutdinov}{Srivastava et~al.}{2014}]{sri14}
Srivastava N.,  Hinton G.,  Krizhevsky A.,  Sutskever I.,   Salakhutdinov R.,
  2014, The journal of machine learning research, 15, 1929

\bibitem[\protect\citeauthoryear{{Stebbins}}{{Stebbins}}{1988}]{Stebbins:1988}
{Stebbins} A.,  1988, \mn@doi [Astrophys. J.] {10.1086/166218}, \href
  {http://adsabs.harvard.edu/abs/1988ApJ...327..584S} {327, 584}

\bibitem[\protect\citeauthoryear{Stebbins \& Veeraraghavan}{Stebbins \&
  Veeraraghavan}{1995}]{Stebbins:1995}
Stebbins A.,  Veeraraghavan S.,  1995, Physical Review D, 51, 1465

\bibitem[\protect\citeauthoryear{Stewart \& Brandenberger}{Stewart \&
  Brandenberger}{2009}]{ste08}
Stewart A.,  Brandenberger R.,  2009, \mn@doi [JCAP]
  {10.1088/1475-7516/2009/02/009}, 0902, 009

\bibitem[\protect\citeauthoryear{Tuntsov \& Pshirkov}{Tuntsov \&
  Pshirkov}{2010}]{tun10}
Tuntsov A.~V.,  Pshirkov M.~S.,  2010, \mn@doi [Phys. Rev.]
  {10.1103/PhysRevD.81.063523}, D81, 063523

\bibitem[\protect\citeauthoryear{Vachaspati \& Vilenkin}{Vachaspati \&
  Vilenkin}{1984}]{Vachaspati:1984dz}
Vachaspati T.,  Vilenkin A.,  1984, \mn@doi [Phys. Rev.]
  {10.1103/PhysRevD.30.2036}, D30, 2036

\bibitem[\protect\citeauthoryear{Vafaei~Sadr \& Movahed}{Vafaei~Sadr \&
  Movahed}{2021}]{vafaei2021clustering}
Vafaei~Sadr A.,  Movahed S.,  2021, Monthly Notices of the Royal Astronomical
  Society, 503, 815

\bibitem[\protect\citeauthoryear{Vafaei~Sadr, Movahed, Farhang, Ringeval  \&
  Bouchet}{Vafaei~Sadr et~al.}{2017}]{vaf17}
Vafaei~Sadr A.,  Movahed S.,  Farhang M.,  Ringeval C.,   Bouchet F.,  2017,
  Monthly Notices of the Royal Astronomical Society, 475, 1010

\bibitem[\protect\citeauthoryear{Vafaei~Sadr, Farhang, Movahed, Bassett  \&
  Kunz}{Vafaei~Sadr et~al.}{2018}]{vaf18}
Vafaei~Sadr A.,  Farhang M.,  Movahed S.,  Bassett B.,   Kunz M.,  2018,
  Monthly Notices of the Royal Astronomical Society, 478, 1132

\bibitem[\protect\citeauthoryear{Vilenkin}{Vilenkin}{1981}]{Vilenkin:1981iu}
Vilenkin A.,  1981, \mn@doi [Phys. Rev. Lett.] {10.1103/PhysRevLett.46.1169,
  10.1103/PhysRevLett.46.1496}, 46, 1169

\bibitem[\protect\citeauthoryear{Vilenkin}{Vilenkin}{1985}]{Vilenkin:1984ib}
Vilenkin A.,  1985, \mn@doi [Phys. Rept.] {10.1016/0370-1573(85)90033-X}, 121,
  263

\bibitem[\protect\citeauthoryear{Vilenkin \& Shellard}{Vilenkin \&
  Shellard}{2000}]{Vilenkin:2000jqa}
Vilenkin A.,  Shellard E. P.~S.,  2000, {Cosmic Strings and Other Topological
  Defects}.
Cambridge University Press, \url
  {http://www.cambridge.org/mw/academic/subjects/physics/theoretical-physics-and-mathematical-physics/cosmic-strings-and-other-topological-defects?format=PB}

\bibitem[\protect\citeauthoryear{Yegnanarayana}{Yegnanarayana}{2009}]{yegnanarayana2009artificial}
Yegnanarayana B.,  2009, Artificial neural networks.
PHI Learning Pvt. Ltd.

\bibitem[\protect\citeauthoryear{Zeldovich}{Zeldovich}{1980}]{Zeldovich:1980gh}
Zeldovich {\relax Ya}.~B.,  1980, Mon. Not. Roy. Astron. Soc., 192, 663

\makeatother
\end{thebibliography}

\bsp	
\label{lastpage}
\end{document}
